\documentclass[pra,twocolumn,showkeywords,bibliography,superscriptaddress,10pt]{revtex4-1}
\usepackage{epsfig}
\usepackage{graphicx}
\usepackage{color}
\usepackage{amsfonts}
\usepackage{amscd}
\usepackage[colorlinks, citecolor=blue, linkcolor=blue,urlcolor=blue]{hyperref}
\usepackage[mathscr]{euscript} 
\usepackage{bookmark} 
\usepackage{amsmath}
\usepackage{epstopdf}
\usepackage{graphicx}
\usepackage{txfonts}
\usepackage{textcomp}

\begin{document}
	
	\title{Robust coherent control in non-Hermitian cavity electromagnonics using counterdiabatic driving}

\author{Guang-Hui Zhang} 

\affiliation{School of Physics \& Optoelectronic Engineering, Anhui University, Hefei
	230601,  People's Republic of China}

\author{Yu-Wen Li}

\affiliation{School of Physics \& Optoelectronic Engineering, Anhui University, Hefei
	230601,  People's Republic of China}

\author{Xue-Ke Song} \email{songxk@ahu.edu.cn}
\affiliation{School of Physics \& Optoelectronic Engineering, Anhui University, Hefei 230601,  People's Republic of China}
\author{Liu Ye}
\affiliation{School of Physics \& Optoelectronic Engineering, Anhui University, Hefei 230601,  People's Republic of China}

\author{Dong Wang} \email{dwang@ahu.edu.cn}
\affiliation{School of Physics \& Optoelectronic Engineering, Anhui University, Hefei
	230601,  People's Republic of China}

\date{\today }

	\begin{abstract}
		
		Here, we propose to use counterdiabatic driving (CD) shortcut and the Floquet engineering to realize the robust and fast state transfer in the dissipation cavity magnon-polaritons non-Hermitian (NH) system.
		For the two-level NH cavity magnon-polaritons Hamiltonian, an accurate and fast population transfer is achieved from the microwave photon to the magnon by two coherent control techniques; counterdiabatic driving shortcut and non-Hermitian shortcuts (NHS). Additionally, by using the CD technique, the population evolution speed of non-Hermitian systems is faster than that via the NHS technique in the broken-$\mathcal{P} \mathcal{T}$-symmetric regime.
		 Furthermore, we compare their performances in the presence of the coupling strength and systematic errors, the CD technique features a broad range of high efficiencies of the transition probability above 99.9\%,
showing that the CD technique is more robustness against these errors than the NHS technique. It is worth noting that this advantage becomes more significant as the gain rate of system parameters increases.
 The work provides a basis for achieving the robust coherent control in NH cavity electromagnonics.

	\end{abstract}
	%\pacs{03.67.Lx, 32.80.Qk, 76.30.Mi}
	
	\maketitle
	
	%\footnotetext[\dag]{Co-first authors}
	%implying that the time evolution of neutrino system should be understood from the point of quantum mechanics
	
	\section{Introduction}
	\label{sec1}
	
	%Open systems with
	%gain and loss naturally exhibit non-Hermitian physics \cite{Ashida}. It has important applications in physical systems, including
	%quantum resonances, superradiance, continuous quantum Zeno effect quantum critical phenomena, Dirac spectra in quantum chromodynamics, and nonunitary
	%conformal field theories.
	%Recently, there has been growing interest in  non-Hermitian physics of cavity magnon-polaritons \cite{DZhang11,BWang,MHarder1}, owing to non-Hermitian Hamiltonians typically describe subsystems of a larger and open system \cite{JGMuga1}.
	%Especially in the context of the PT-symmetric non-Hermitian systems \cite{CarlMBender111,AMostafazadeh,CMBender1,JWWen,XGLi}, one case of PT-symmetric and unbroken, where the eigenvalues of the PT-symmetric Hamiltonian function are true, and the other case of PT-symmetric and broken, the eigenvalues are the complex conjugate pairs, and a PT-symmetric Hamiltonian evolves faster than Hermitian in a two-state
	%quantum system \cite{CarlMBender111}.
	%Recently, there has been growing interest in  non-Hermitian physics of cavity magnon-polaritons \cite{DZhang11,BWang,MHarder}, owing to non-Hermitian Hamiltonians typically describe subsystems of a larger and open system \cite{JGMuga1}. It has important applications in physical systems, including
	%quantum resonances, superradiance, continuous quantum Zeno effect quantum critical phenomena, Dirac spectra in quantum chromodynamics, and nonunitary
	%conformal field theories.

	In recent years, the studies of non-Hermitian (NH) systems have attracted considerable attention. Especially,
	a large class of NH Hamiltonians satisfying parity-time symmetry can exhibit entirely real
	spectra \cite{PhysRevLett.80.5243,MUGA2004357}. It implies that there may exist the rich and exotic physical properties in
	the vicinity of the spectral transition in parity-time symmetric NH systems \cite{ypd8-r9gq,PhysRevLett.98.040403,BENDER20101616,PhysRevA.99.062122,Peng2014}. It has wide applications in many prominent phenomena and subjects, including
	quantum resonances \cite{AIMagunov1999}, continuous quantum Zeno effect \cite{PhysRevA.101.052122}, and quantum
	critical phenomena \cite{Chang2020}, etc.  The non-Hermitian descriptions of
	physical reality are found in a wide variety of classical and quantum systems, such as the
	gain and loss photons in photonics \cite{PhysRevA.101.052122}, dissipation in open quantum systems, and so on. In particular, when the dissipation of the cavity
	magnon-polaritons is considered, the system may be described by the NH Hamiltonian \cite{Zhang2017,PhysRevB.95.214411,PhysRevB.94.224410}. As a research hotspot, non-Hermitian systems exhibit unique physical properties that distinguish them from traditional Hermitian systems \cite{PhysRevResearch.6.023135,Zhang2025,Dogra2021,article1}. For instance, under periodic boundary conditions, the eigenstates of NH Hamiltonian systems exhibit extended Bloch waves. Under open boundary conditions, a skin effect that is localized to the boundary and decays exponentially occurs, which is not the Hermitian skin effect \cite{Li2024,PhysRevX.13.021007,Zhou2023}.
	When the system parameters are changed, the $\mathcal{P} \mathcal{T}$ -symmetric condition is broken and a phase transition from the $\mathcal{P} \mathcal{T}$ -symmetric regime to the broken-$\mathcal{P} \mathcal{T}$-symmetric regime can occur, the stability of the system changes accordingly, which in turn affects the dynamic behavior of the system \cite{PhysRevA.95.013843}. And the dividing points between the two phases are the exceptional points (EPs) unique to NH Hamiltonian systems, where the eigenstates and eigenvalues of the system are simultaneously degenerate \cite{Li2023,Ding2022,PhysRevLett.119.190401,doi:10.1126/science.aaw8205}.
	
	%Much effort are made
	%to explore such interesting systems in both theory and experiment in different physical systems. For example, in 2014, Peng \emph{et al.} \cite{Peng2014}
	%observed the non-reciprocity of the PT-symmetry-breaking phase in whispering-gallery microcavities.
	%In 2017, Kawabata \emph{et al.} \cite{PhysRevLett.119.190401}
%	studied the information retrieval and criticality in parity-time symmetric systems.
	%In 2019, Wu \emph{et al.} \cite{doi:10.1126/science.aaw8205} experimentally demonstrated the parity-time symmetry
	%breaking in a single nitrogen-vacancy center in diamond.

	Cavity electromagnonics \cite{PhysRevLett.113.156401,Zhang2015,doi:10.1126/sciadv.1501286,10.1063/5.0046202}, a hybridization of magnon and
	cavity photon, provides a useful platform for the flexibility and controllability
	of the hybridized states between spin and photon. It is of great significance in studying the
	potential applications in light-matter interaction and quantum
	information technology \cite{PhysRevLett.113.083603,PhysRevLett.123.107701,PhysRevA.106.012609,PhysRevB.107.054413}.
	 Floquet cavity electromagnonics was proposed by introducing Floquet engineering into cavity magnon-polariton \cite{PhysRevLett.125.237201,wc1j-mq69,PhysRevLett.134.063602,GRIFONI1998229}. Due to spin excitations in single-crystal and highly purified YIG, magnons exhibit a very low dissipation rate, making them excellent carriers for quantum information \cite{PhysRevA.106.012609}. A new coupling regime of Floquet ultrastrong coupling was observed previously, which provides a new approach to achieving more efficient quantum modulation \cite{PhysRevLett.125.237201}.
	Recently, Floquet engineering is shown to be an effective approach for
	controlling the dynamics of NH cavity electromagnonics systems \cite{Zhang2017,PMID:30517807,Claeys:2019mnm,PhysRevLett.125.237201,PhysRevLett.128.233201,PhysRevA.106.033711}.
	It is shown that the high-fidelity coherent control of quantum
	state in cavity electromagnonics is realized by the Floquet engineering.
	In 2020, Xu \emph{et al.} \cite{PhysRevLett.125.237201}
	used  Floquet drive of the periodic temporal modulation to enable the manipulation of the interaction between
	hybridized cavity electromagnonic modes.
	In 2022, Jiang \emph{et al.} \cite{PhysRevLett.128.233201} presented the theoretical and experimental
	demonstration of quantum amplification on periodically driven spins by Floquet engineering.

	On the other hand, there are many coherent control methods \cite{ZhenKunWu2013228,YiXuanWu202242507,JiaNingZhang202351303,QiPingSu202161501,7h6g-1z57,PhysRevA.111.012406} that can be employed to investigate the quantum dynamics and
	manipulation of quantum states in NH systems, such as adiabatic passages \cite{PhysRevA.89.033403,JiaNingZhang2025014206}, shortcuts
	to adiabaticity (STAs) \cite{PhysRevA.84.023415,PhysRevB.111.064301,PhysRevA.105.013714,Zhang_2021,Song_2023,PhysRevA.105.053710}, machine learning \cite{PhysRevLett.126.240402,PhysRevA.103.012419}, and so on. Among them, STA technology shows the unique advantages in controlling the evolution of quantum systems in a fast and accurate manners. STA technology is an effective method for suppressing nonadiabatic transitions. Its basic idea  is to add counteradiabatic terms to the original time-dependent Hamiltonian to achieve the suppression of transitions between instantaneous eigenstates \cite{UNANYAN199748,PhysRevB.103.075118,m1cc-l6ng,Tancara2025,Wu_2021,Du2016}.
	The different approaches to engineer STA, including counter-adiabatic driving (CD) \cite{Berry_2009,PhysRevA.111.L031301,PhysRevLett.105.123003,PhysRevResearch.6.033332,PhysRevB.109.245421,PhysRevA.93.052324,PhysRevLett.126.023602}, inverse engineering based on Lewis-Riesenfeld invariants \cite{PhysRevA.111.012623,PhysRevA.109.062610,PhysRevB.110.155104,Chen2010,Song:21}, are proposed to accelerate a quantum adiabatic process to reproduce the same final population.  The CD shortcut technology is a common method for realizing shortcuts
		to adiabaticity, which can address the limitations of slow evolution speed and the occurrence of non-adiabatic transitions in traditional adiabatic approaches, by adding an auxiliary Hamiltonian \cite{PhysRevA.94.063411,PhysRevA.105.022614}. By designing a specific driving Hamiltonian, the quantum system can precisely evolve along the instantaneous eigenstates of the initial Hamiltonian, thereby shortening the evolution time and avoiding the loss of population transfer efficiency caused by non-adiabatic transitions \cite{PhysRevB.111.064301,PhysRevB.110.024304,PhysRevA.111.L031301}. The STA is originally studied in Hermitian systems and subsequently has been extended to the two-level and  three-level NH systems in recent years \cite{PhysRevA.87.052502,PhysRevA.89.063412,PhysRevA.111.022410,PhysRevA.93.052109,PhysRevLett.122.050404,PhysRevA.105.013714,ZHANG2023106421}.
	For example, in 2022, Luan \emph{et al.} \cite{PhysRevA.105.013714} used CD shortcut mehtod to realize a population transfer in general two-level NH systems. In 2023, Zhang \emph{et al.} \cite{ZHANG2023106421}
	studied how to create ultracold deeply-bound molecules by NH stimulated Raman shortcut-to-adiabatic passage.
	
	%scalablehas a significantly important purpose to prepare a desired
	%state as fast as possible, with a high fidelity in

	In this paper, we use the shortcuts to adiabaticity to realize the robust and fast state transfer in cavity electromagnonics, based on the CD and the Floquet engineering. 
	For the two-level NH Hamiltonian, the CD technique can achieve an accurate relative population transfer between the microwave photons and magnons. Specifically,  we realize perfect state transfer in the two level NH Hamiltonian system using two quantum control techniques NHS and CD. We find that the evolution of the populations in the NH Hamiltonian system is faster when using the CD technique.
	Furthermore, we analyze the effects of the coupling strength  and the systematic errors on the fidelities of the population transfer for two coherent control techniques NHS and CD, respectively, when the NH Hamiltonian system is in the broken-$\mathcal{P} \mathcal{T}$-symmetric regime. By precisely suppressing non-adiabatic transitions, CD technique enables faster state transitions than the NHS, while maintaining higher fidelity and stronger robustness. We introduce the Floquet engineering into the non-Hermitian cavity magneto-vibration system. By periodically driving and dynamically modulating the frequency of the magnon, the energy level difference of the system exhibits controllable time evolution characteristics. On this basis, we combine the two techniques of the CD and the NHS, respectively. CD precisely offsets non-adiabatic coupling by supplementing the Hamiltonian, and NHS compensates for non-adiabatic loss through the imaginary term of the diagonal element of the Hamiltonian. Ultimately, it achieves more efficient quantum state transfer and significantly broadens the applicable parameter range and anti-error capability of the two control methods.
	We conclude that the CD technique is more robust against these errors, with ultrahigh fidelity of transition probability. This offers an effective approach for achieving high-fidelity coherent control of two-level NH quantum systems in cavity electromagnonics.

	This paper is organized as follows: In Sec. \ref{sec2}, we introduce the model and the total Hamiltonian for the NH cavity-magnon system.
	In Sec. \ref{sec3}, we provide a brief description of two coherent control techniques, including  NHS and CD. We discuss coupling strength
	and systematic errors for the above coherent control techniques
	in Sec. \ref{sec4}.
	A summary is provided in Sec. \ref{sec5}.

	\section{MODEL AND HAMILTONIAN}
	\label{sec2}
\begin{figure}[t]%[tpb]  % picture 1
		\begin{center}
			\includegraphics[width=8 cm,angle=0]{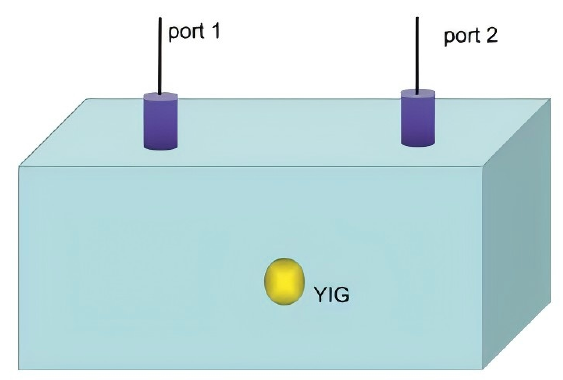}
			\caption{The yttrium iron garnet (YIG) sphere is coupled to a
				microwave cavity. }\label{mode}
		\end{center}
	\end{figure}

	We consider the model that a small yttrium iron garnet (YIG)
	sphere with a diameter  is mounted in a 3D rectangular
	microwave cavity in the Fig. \ref{mode}, the two ports 1 and 2 are both for measurement and feeding microwave fields into the cavity.
	The system of the
	cavity-magnon is described by the Hamiltonian \cite{PhysRevLett.113.156401,PhysRevA.106.033711}:
	\begin{align}
		H_{s}/\hbar=\omega_{c} a^{\dagger} a+\omega_{m} m^{\dagger} m+g_{m}\left(a+a^{\dagger}\right)\left(m^{\dagger}+m\right),
		\label{Hamiltoniancavity}
	\end{align}
	where $ a^{\dagger}( a)$ is the creation (annihilation) operator of the
	cavity with frequency $\omega_{c}$, and $\omega_{m}$ is the frequency of the magnon mode. The coupling strength $g_{m}$ between the cavity and the magnon is $g_{m}=\omega \gamma \sqrt{\omega_{c} \mu_{0} N s / 2 V_{a}}$ in the YIG sphere, where $\omega$ describes the spatial overlap and polarization matching conditions between the microwave field and the magnon, $V_{a}$ is the mode
	volume of the microwave cavity resonance, $\mu_{0}$ is the
	vacuum permeability, $N$ is the total number of spins
	and $s$ is the spin number of the ground state \cite{PhysRevLett.113.156401}.
	 When the coupling strength satisfies this condition $g_{m} \ll\left\{\omega_{c}, \omega_{m}\right\}$, the interaction term in the Hamiltonian  is transformed into $g_{\mathrm{m}}\left(a m^{\dagger}+a^{\dagger} m\right)$ under the rotating-wave
	 approximation.
	
	The quantum Langevin equations describing the system are
		given by \cite{RevModPhys.86.1391}
	%%%%%%%%%%%%%%%%%%%%%%%%%%%%%%%%%%%%%%%%%
	\begin{eqnarray}
		\dot{a}(t)  &= -i\left[a, H_{\mathrm{s}}\right]-\left(\kappa_{\mathrm{int}}+\sum_{i=1}^{2} \kappa_{i}\right) a \nonumber\\
		&+\sqrt{2 \kappa_{\mathrm{int}}} a_{\mathrm{int}}^{(\mathrm{in})}+\sum_{i=1}^{2} \sqrt{2 \kappa_{i}} a_{i}^{(\mathrm{in})},\\
		\dot{m}(t)  &= -i\left[m, H_{\mathrm{s}}\right]-\kappa_{\mathrm{m}} m+\sqrt{2 \kappa_{\mathrm{m}}} m^{(\mathrm{in})},
	\end{eqnarray}
	where $\kappa_{int}$ is the intrinsic loss rate of the cavity mode, $\kappa_{i}$ (i = 1, 2) is the decay rate of the cavity mode,
	$a_{i}^{(\mathrm{in})}$ (i = 1, 2) and $m^{(\mathrm{in})}$ are input noise operators.
	There exists a relation between input and output fields to the intra-cavity field \cite{Zhang2017}
	\begin{eqnarray}
		a_{\mathrm{int}}^{\text {(out })}+a_{\mathrm{int}}^{(\mathrm{in})}&=&\sqrt{2 \kappa_{\mathrm{int}}} a, \\
		a_{i}^{\text {(out) }}(t)+a_{i}^{(\mathrm{in})}&=&\sqrt{2 \kappa_{i}} a \quad(i=1,2),
	\end{eqnarray}
	%%%%%%%%%%%%%%%%%%%%%%%%%%%%%%%%%%%%%%%%%%
	where $a_{i}^{(\mathrm{out})}$ (i = 1, 2, int) is the output noise operator. If no input field is related to the intrinsic loss of the intracavity field,
	the input noise operator $a_{\mathrm{int}}^{(\mathrm{in})}$ is zero. We consider the
	perfect field-feeding case, the output noise operators of the
	cavity via ports 1 and 2 are $a_{i}^{\text {(out) }}=0$ $(i=1,2)$. The relationships between the input and output operators are reduced to
	$a_{1}^{(\mathrm{in})}=\sqrt{2 \kappa_{1}} a(t), a_{2}^{(\mathrm{in})}=\sqrt{2 \kappa_{2}} a$, and $a_{int}^{(\mathrm{out})}=\sqrt{2 \kappa_{int}} a$ \cite{PhysRevA.106.012609}.
	Also, if there is no input field the
	of intrinsic damping rate of the magnon mode, we have $m^{in}=0$.
	In the case, the quantum Langevin equations are simplified as
	%%%%%%%%%%%%%%%%%%%%%%%%%%%
	\begin{eqnarray}
		\dot{a} &=& -i\left[\omega_{c}+i \kappa_{c} \right] a-i g_{m} m, \nonumber\\
		\dot{m} &=& -i\left(\omega_{m}-i \kappa_{m}\right) m-i g_{m} a,
		\label{laneff}
	\end{eqnarray}
	%%%%%%%%%%%%%%%%%%%%%%%%%%%
	where $\kappa_{c}=\left(\kappa_{1}+\kappa_{2}-\kappa_{i n t}\right)$ is the total damping rate of the
	cavity mode.
	
	On the other hand, the equations of motion in Eq. (\ref{laneff}) can be rewritten as
	%%%%%%%%%%%%%%%%%%%%%%%%%%%
	\begin{align}
		\dot{a}  =  -i\left[a, H_{\mathrm{eff}}\right], \quad \dot{m}  =  -i\left[m, H_{\mathrm{eff}}\right],
	\end{align}
	%%%%%%%%%%%%%%%%%%%%%%%%%%%
	where $H_{\mathrm{eff}}$ is the effective NH Hamiltonian of the
	cavity-magnon system as
	%%%%%%%%%%%%%%%%%%%%%%%%%%%
	\begin{align}
		H_{\mathrm{eff}}=\left(\omega_{c}+i \kappa_{c}\right) a^{\dagger} a+\left(\omega_{m}-i \kappa_{m}\right) m^{\dagger} m+g_{m}\left(a m^{\dagger}+a^{\dagger} m\right).
		\label{noHamiltoniancavity1}
	\end{align}
	%%%%%%%%%%%%%%%%%%%%%%%%%%%
		\begin{figure}[t]%[tpb]  % picture 1
		\begin{center}
			\includegraphics[width=8 cm,angle=0]{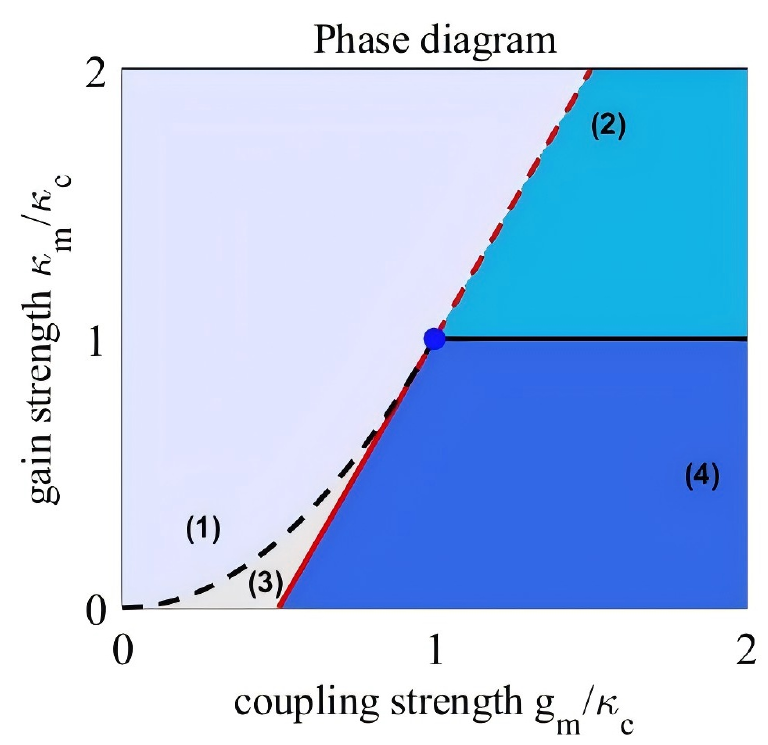}
			\caption{Phase diagram under different conditions of the gain rate $\kappa_{m}$ and the coupling strength $g_{m}$ in units of the cavity decay rate $\kappa_{c}$. There are two borders, the red line shows the border between the $\mathcal{P} \mathcal{T}$-symmetric phase (on the right hand) and the broken-$\mathcal{P} \mathcal{T}$-symmetric phase ( on the left hand). The black dashed curve and black solid line are the border between the asymptotically stable (below the border) phase and the unstable (above the border) phase. }\label{fig11}
		\end{center}
	\end{figure}	
	Furthermore, we use the tools of Floquet engineering to realize a high-fidelity control of the quantum states \cite{Claeys:2019mnm}. To do this, a sinusoidal frequency modulation of the magnon mode and the cavity mode is applied, and the corresponding Floquet Hamiltonian is expressed as
	%%%%%%%%%%%%%%%%%%%%%%%%%%%
	\begin{align}
		H_{F}=\varepsilon_{m} m^{\dagger} m \cos \left(\omega_{d} t\right),
		\label{noHamiltoniancavity2}
	\end{align}
	where $\varepsilon_{m}$ and $\omega_{d}$ are the strength and the frequency of the
	driving field, respectively. In the Heisenberg picture, the mode operators follow the dynamical equation ${\dot{\textbf{A}}(t)}=-i{H}{\textbf{A}}(t)$. In the subspace of the single excitation, the total Hamiltonian can be written as a coefficient matrix
	%%%%%%%%%%%%%%%%%%%%%%%%%%%
	\begin{align}
		H=\left(\begin{array}{cc}
			\omega_{c}+i \kappa_{c} & g_{m} \\
			g_{m} & \omega_{m}(t)-i \kappa_{m}
		\end{array}\right) \label{H},
	\end{align}
	where $H=H_{\mathrm{eff}}+H_{F}$ with $\omega_{m}(t)=\omega_{m}+\varepsilon_{m} \cos \left(\omega_{d} t\right)$, $\varepsilon_{m}$ and $\omega_{d}$ are the strength and frequency of the
	driving field. In this sense, the Floquet driving can be realized through
	the total frequency modulation of the magnon mode, and thus we can obtain a periodically adjustable non-Hermitian system.
	
		We consider the scenario where $\mathcal{P} \mathcal{T}$ symmetry between the cavity decay rate $\kappa _{c}  $ and the mechanical gain strength  is not satisfied, implyin that only the relation  $\Delta =\omega  _{c}-\omega  _{m} (t)= \omega _{1} $ is always fulfilled. The Hamiltonian is rewritten as
	\begin{align}
		\hat{H}=\hbar \begin{pmatrix}
			\hat{a}^{\dagger }  &\hat{b}^{\dagger }
		\end{pmatrix}\begin{pmatrix}
			\omega _{1}-i\kappa _{c}  &-g_{m}  \\
			-g_{m} &\omega _{1}+i\kappa _{m} 
		\end{pmatrix}\begin{pmatrix}
			\hat{a}\\\hat{b}
		\end{pmatrix}
	\end{align}
where $\hat{a}$ ($\hat{a^{\dagger } } $) and $\hat{b}$ ($\hat{b^{\dagger } } $) are the annihilation (creation) operators of the cavity field and the mechanical oscillator,
respectively; and $g_{m}$  is the coupling strength. By diagonalizing the matrix, the eigenfrequencies of the supermodes $\hat{A} _{\pm } =\left ( \hat{a} \pm \hat{b}  \right )$ can be obtained as
	\begin{align}
		\omega _{\pm } =\omega _{1}-\frac{i}{2}\left ( \kappa _{c} -\kappa _{m}  \right ) 
		\pm \sqrt{g _{m}^{2}-\frac{1}{4}\left ( \kappa _{c} +\kappa _{m}  \right )^{2}    }   
	\end{align}
	%%%%%%%%%%%%%%%%%%%%%%%%%%%

In Fig. \ref{fig11}, the phase diagram under different conditions of the gain rate $\kappa_{m}$ and the coupling strength $g_{m}$ in units of the cavity decay rate $\kappa_{c}$. The system is in an unstable state, when the parameters of the phase diagram regions (1) and (2) satisfy $\kappa_{c}> \kappa_{m}$. And the system is in an asymptotically stable state, when the parameter conditions in regions (3) and (4) of the phase diagram are $\kappa_{c}<\kappa_{m}$. When $g _{m}> \left ( \kappa _{c} +\kappa _{m}  \right ) /2$, the system is in the $\mathcal{P} \mathcal{T}$ symmetry, where the eigenvalues have the same imaginary part and two differen real parts. This case corresponds to regions (2) and (4) in the phase diagram.  
	When $g _{m}< \left ( \kappa _{c} +\kappa _{m}  \right ) /2$, the eigenvalues of the Hamiltonian only have an identical real part and two different imaginary parts in the broken-$\mathcal{P} \mathcal{T}$-symmetric regime, which is described by the regimes
	(1) and (3) in the phase diagram shown in Fig. \ref{fig11}.
And the border point between the $\mathcal{P} \mathcal{T}$ symmetry and broken-$\mathcal{P} \mathcal{T}$ symmetry satisfies the relation $g _{m}= \left ( \kappa _{c} +\kappa _{m}  \right ) /2$ and is defined as an exceptional point (EP) as shown by the red line and blue point in the phase diagram \cite{PhysRevA.104.053518}.

	\section{ COHERENT CONTROL TECHNIQUES}
	\label{sec3}
	
	\subsection{\texorpdfstring{Non-Hermitian shortcuts }{Non-Hermitian shortcuts}}
				
	NHS provides a method to cancel the nonadiabatic losses by adding an imaginary term in the diagonal elements of the Hamiltonian, so that an arbitrarily fast population transfer is realized \cite{PhysRevA.87.052502,PhysRevA.89.063412}. In the whole process, there is no need to increase the coupling.
					
					When the dissipation of the system is not be considered, i.e., $\kappa_{c}=\kappa_{m}=0$, we can get the eigenstates (adiabatic states) of the Eq. (\ref{H}) as
					\begin{align}
						\left| {{\varphi _ + }(t)} \right\rangle & = \cos \theta (t)\left| 0 \right\rangle  + \sin \theta (t)\left| 1 \right\rangle ,\nonumber\\
						\left| {{\varphi _ - }(t)} \right\rangle  &=  - \sin \theta (t)\left| 0 \right\rangle  + \cos \theta (t)\left| 1 \right\rangle,
					\end{align}
					with eigenvalues ${\lambda _ {1,2} } = [{\omega _c} + {\omega _m}(t)]/2 \pm \sqrt {g_m^2 + {\Delta ^2}/4} $, the mixing angle $\theta(t)=\frac{1}{2} \arctan \left[2 g_{m} / \Delta(t)\right]$, and $\Delta=\omega_{c}-\omega_{m}(t)$.
					A simple population transfer is completely realized by the fast adiabatic passage when the system evolves adiabatically along the adiabatic state $\left| {{\varphi _ +}(t)} \right\rangle$ or $\left| {{\varphi _ - }(t)} \right\rangle$.

				The adiabatic basis
					${\left| {{\varphi _ + }(t)} \right\rangle , \left| {{\varphi _ + }(t)} \right\rangle} $ is connected with the bare basis ${\left| {0} \right\rangle , \left| {1} \right\rangle} $
					via a rotation matrix
					%%%%%%%%%%%%%%%%%%%%%%%%%%%%%%%
					\begin{eqnarray}
						R(\theta ) = \left( {\begin{array}{*{20}{c}}
								{\cos \theta (t)}&{\sin \theta (t)} \\
								{ - \sin \theta (t)}&{\cos \theta (t)}
						\end{array}} \right).
					\end{eqnarray}
					\begin{figure}[t]%[tpb]  % picture 1
						\begin{center}
							\includegraphics[width=8.5 cm,angle=0]{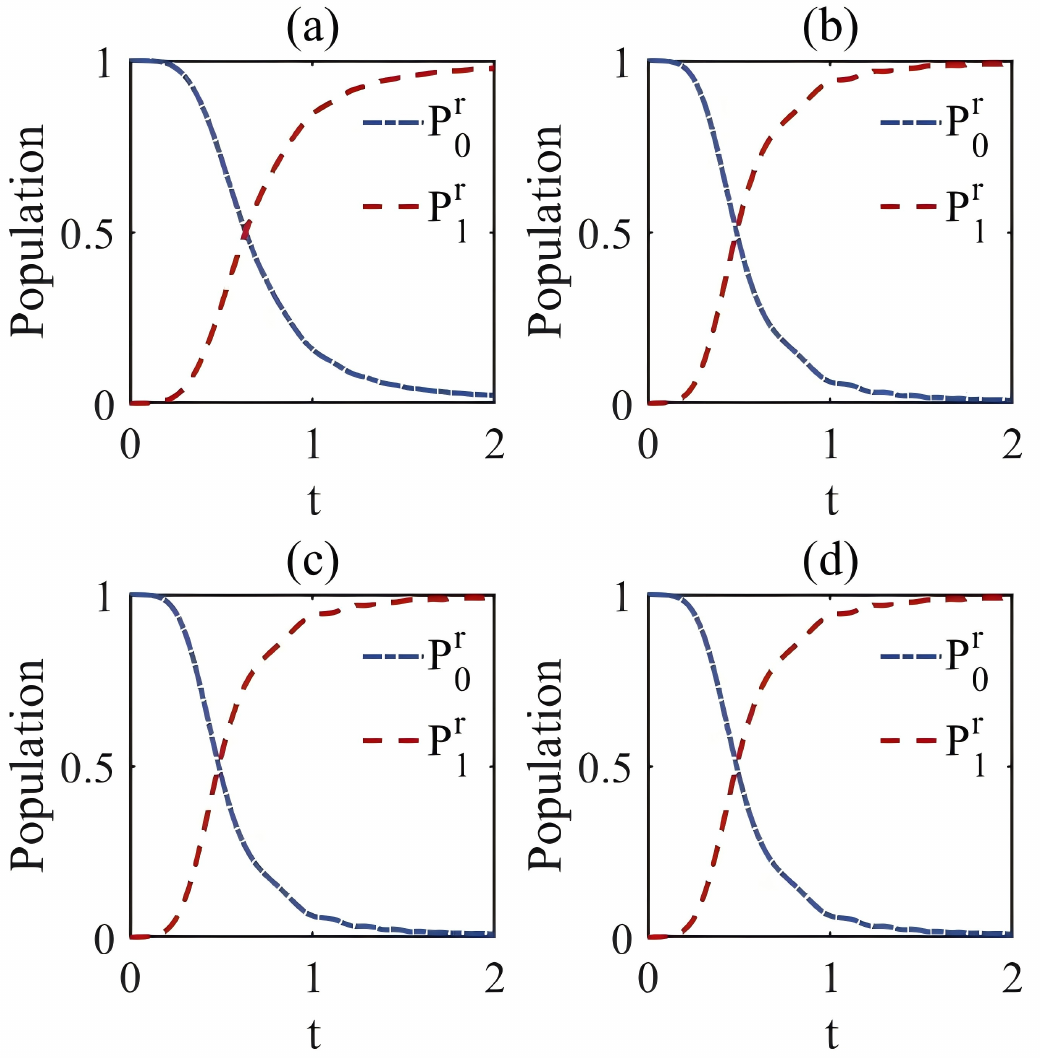}
							\caption{The figures  plot the time evolution of the relative populations as a function of time  with parameters $\omega_{c}/\omega_{d}=85, $ $\omega_{m}/\omega_{d}=35$ and $\varepsilon_{m}/\omega_{d}=50$, $P^{r}_{0}$ and $P^{r}_{1}$ are the relative
								populations of the microwave photons and the magnons, respectively.  (a) $g_{m} /\omega_{d}=0.1$, (b) $g_{m} /\omega_{d}=0.3$, (c) $g_{m} /\omega_{d}=0.6$, (d) $g_{m} /\omega_{d}=1$. }\label{fig1}
						\end{center}
					\end{figure}		
					In the adiabatic basis, the Hamiltonian $H$ without the dissipation is expressed as
					\begin{align}{H_a} = \left( {\begin{array}{*{20}{c}}
								{{\lambda _ 2 }(t)}&{ - i\dot \theta (t)} \\
								{i\dot \theta (t)}&{{\lambda _ 1 }(t)}
						\end{array}} \right).
					\end{align}
					If the dissipation of the system is nonzero ($\kappa_{c} \ne \kappa_{m} \ne 0$), in the adiabatic basis $\left| {{\varphi_ \pm }(t)} \right\rangle $, the Eq. (\ref{H}) can be rewritten
					as
					
					\begin{align}
						{H}'=\begin{pmatrix}
							E_{- - } &E_{- +  }  \\
							E_{+  - } &E_{+  +  } 
						\end{pmatrix} \label{eq.16}
					\end{align} 
					where
					\begin{align}
						E_{--}=\lambda _{2}\left ( t \right )  +i \kappa_{c} \cos ^{2} \theta(t)-i \gamma_{m} \sin ^{2} \theta(t)\tag{16a} \\
						E_{-+}=-i\left(\kappa_{c}+\gamma_{m}\right) \sin \theta(t) \cos \theta(t)-i \dot{\theta}(t)\tag{16b} \\
						E_{+-}=-i\left(\kappa_{c}+\gamma_{m}\right) \sin \theta(t) \cos \theta(t)+i \dot{\theta}(t) \tag{16c}\\
						E_{++}=\lambda _{2}\left ( t \right )+i \kappa_{c} \sin ^{2} \theta(t)-i \gamma_{m} \cos ^{2} \theta(t)\tag{16d}
					\end{align}
					Under the adiabatic basis, the Hamiltonian Eq. (\ref{eq.16}) contains non-diagonal matrix elements, which correspond to the non-adiabatic coupling effect and will cause the evolution of the system state to deviate from the target state. To eliminate non-adiabatic coupling effects, let $-\frac{1}{2}i({\kappa _c} +{\kappa _m})\sin2\theta (t) - i \dot{\theta} (t)=0$, one can reach the goal of the quantum state
					transfer from the cavity mode to the magnon mode along the $\left| {{\varphi _ +}(t)} \right\rangle$
					at any arbitrary speed.That is, the state $\left| {{\varphi _ + }(t)} \right\rangle $ initially coincides with state $\left| 0\right\rangle $
					and finally with state $\left| 1 \right\rangle$.

					For simplicity, we consider that the dissipation of the system
					satisfies the condition $\kappa_{c}=\kappa_{m}=\kappa$, the Hamiltonian $H'$ becomes
					%%%%%%%%%%%%%%%%%%%%%%%%%%%%%%%%%%%%%%%
					\begin{align}
						H'=\left(\begin{array}{cc}
							\lambda_{2}(t)+i \kappa \cos 2 \theta(t) & 0 \\
							2 i \dot{\theta}(t) & \lambda_{1}(t)-i \kappa \cos 2 \theta(t)
						\end{array}\right),
					\end{align}
					and the nonadiabatic coupling term is obtained as
					\begin{align}
						\dot{\theta}(t)=\frac{g_{m}}{\Delta^{2}(t)+4 g_{m}^{2}} \dot{\Delta}(t),
					\end{align}
					where  $\dot{\Delta}(t)=\varepsilon_{m} \omega_{d} \sin \left(\omega_{d} t\right)$.
					To implement the population transfer, we choose the dissipation function as
					\begin{align}
						\kappa=-\frac{1}{2} \frac{\dot{\Delta}(t)}{\sqrt{\Delta^{2}(t)+4 g_{m}^{2}}}.
					\end{align}
					The relative
					populations of the NH system are shown in Fig. \ref{fig1}, we plot the evolution of $P_{0}^{r}$ and $P_{1}^{r}$ , which describe the populations of the microwave photon and the magnon, respectively. It can be observed that a relative population transfer is achieved from the level $\left| 0 \right\rangle$ to the level $\left|1\right\rangle$. Specifically, over the time of system evolution  from $t=0$ to $t=2$, the population of level $\left|0\right\rangle$ gradually decreases, while that of level $\left|2\right\rangle$ correspondingly increases.As the parameter $g_{m}/\omega_{d}$ increases, it can be observed that the population $P_{r}^{1}$ improves at the final time $t=2$, increasing from approximately 97.6$\%$ in Fig. \ref{fig1}(a) to about 99$\%$ in Fig. \ref{fig1}(d). We can find that the NHS protocol highly transfers the population number of the photonic state $\left | 0  \right \rangle$ to be adiabatically transferred to the magnonic state $\left | 1  \right \rangle$ under both strong and weak coupling conditions.

				\subsection{\texorpdfstring{Counterdiabatic driving}{Counterdiabatic driving}}
					
					\begin{figure}[b]%[tpb]  % picture 1
						\begin{center}
							\includegraphics[width=8.5 cm,angle=0]{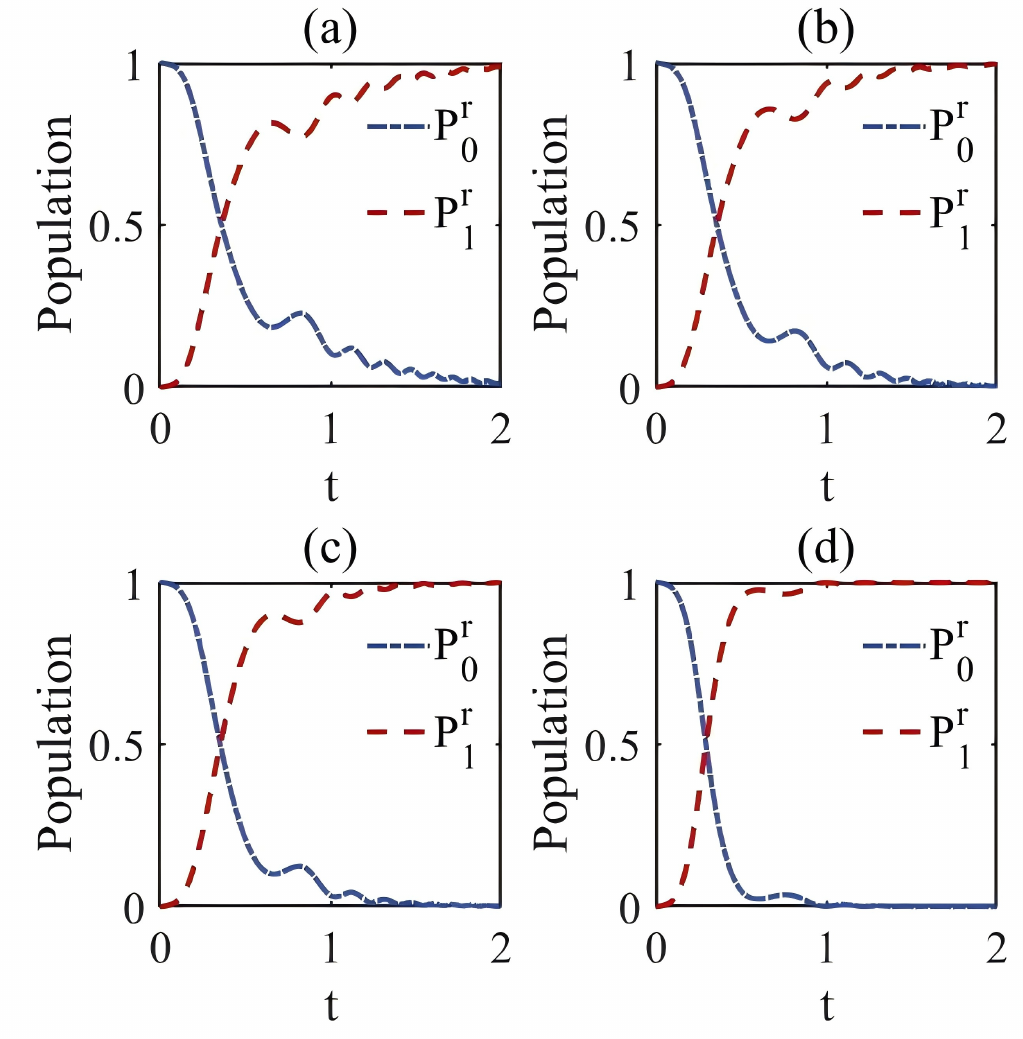}
							\caption{The figures  plot the time evolution of the relative populations as a function of time  with parameters $\omega_{c}/\omega_{d}=85, $ $\omega_{m}/\omega_{d}=35$ , $\varepsilon_{m}/\omega_{d}=50$, $g_{m}=1$ and $P^{r}_{0}$ and $P^{r}_{1}$ are the relative
								populations of the microwave photons and the magnons, respectively.  (a) $\kappa _{c}=1,\kappa _{m}=0.3$, (b) $\kappa _{c}=1,\kappa _{m}=0.6$, (c) $\kappa _{c}=1,\kappa _{m}=1$, (d) $\kappa _{c}=2, \kappa _{m}=2$.}\label{fig2}
							\end{center}
					\end{figure}
					Counteradiabatic driving achieves the shortcut by adding the supplementary Hamiltonian $H_c$ to the reference Hamiltonian $H$. In this case, the nonadiabatic coupling is cancelled out and the
					dynamics follows exactly the approximate adiabatic evolution driven by the Hamiltonian $H$ \cite{PhysRevA.84.023415,PhysRevLett.105.123003}.
					For the Hamiltonian in Eq. (\ref{H}), its eigenvalues are
					\begin{align}
						{\lambda _ {1,2}(t) } &= [{\omega _c} + i{k_c} + {\omega _m}(t) - i{k_m}]/2 \pm \sqrt {g_m^2 + {{\Delta^{'}}^2}/4},
					\end{align}
					and the corresponding eigenstates are
					\begin{align}
						\begin{array}{*{20}{l}}
							{\left| {{\phi _ + }(t)} \right\rangle }&{ = \left( {\begin{array}{*{20}{c}}
										{{g_m}} \\
										{{A_ + }(t)}
								\end{array}} \right)/{N_ + {{(t)}}},} \\
							{\left| {{\phi _ - }(t)} \right\rangle }&{ = \left( {\begin{array}{*{20}{c}}
										{{g_m}} \\
										{{A_ - }(t)}
								\end{array}} \right)}/{N_ - }(t)
						\end{array}
					\end{align}
					with $\Delta^{'}=\omega _{m}(t) - i{k_m} - {\omega _c} - i{k_c}$, ${A_ \pm }(t) = \Delta^{'}/2\pm \sqrt {g_m^2 + {{\Delta^{'}}^2}/4}  $,  and ${N_ \pm }(t) = \sqrt {{g_m}^2 + {A_ \pm }^2(t)}  \hfill $.
					The adjoint Hamiltonian of $H$ is
					\begin{align}
						H^\dag=\left(\begin{array}{cc}
							\omega_{c}-i \kappa_{c} & g_{m} \\
							g_{m} & \omega_{m}(t)+i \kappa_{m}
						\end{array}\right) \label{HT1},
					\end{align}
					with eigenvalues $\lambda^\ast _ {1,2}(t) $ and normalized eigenstates are $\left| {{\hat{\phi}  _ \pm  }(t)} \right\rangle$.
					The Hamiltonian $H_c(t)$ of the external driving field is written as
					\begin{align}
						{H_c}(t) = i\hbar \left[ \begin{gathered}
							\frac{{\left| {{\phi _ + }(t)} \right\rangle \left\langle {{{\hat \phi }_ + }(t)} \right|{\partial _t}H\left| {{\phi _ - }(t)} \right\rangle \left\langle {{{\hat \phi }_ - }(t)} \right|}}{{{\lambda _2}(t) - {\lambda _1}(t)}} \hfill \\
							+ \frac{{\left| {{\phi _ - }(t)} \right\rangle \left\langle {{{\hat \phi }_ - }(t)} \right|{\partial _t}H\left| {{\phi _ + }(t)} \right\rangle \left\langle {{{\hat \phi }_ + }(t)} \right|}}{{{\lambda _1}(t) - {\lambda _2}(t)}} \hfill \\
						\end{gathered}  \right],
					\end{align}
					where \begin{align}
						\begin{array}{*{20}{l}}\vspace{2ex}
							{\left\langle {{{\hat \phi }_ + }(t)} \right|}&{ = \left( {{g_m}\,\,{\text{     }}{A_ + }(t)} \right)/{N_ + }(t),} \\
							{\left\langle {{{\hat \phi }_ - }(t)} \right|}&{ = \left( {{g_m}\,\,{\text{     }}{A_ - }(t)} \right)/{N_ - }(t).}
						\end{array}
					\end{align}
					This gives
					\begin{align}
						H_{c}(t)=\hbar\left(\begin{array}{cc}
							0 & Q(t) \\
							-Q(t) & 0
						\end{array}\right),
					\end{align}
					where
					\begin{align}
						Q(t)=\frac{ig_{m} w_{d}  \varepsilon_{m} sin(w_{d}t)}{{\Delta^{'}}^2+4g_m^2}.
					\end{align}
					\begin{figure*}[t]%[tpb]  % picture 1
						\begin{center}
							\includegraphics[width=15 cm,angle=0]{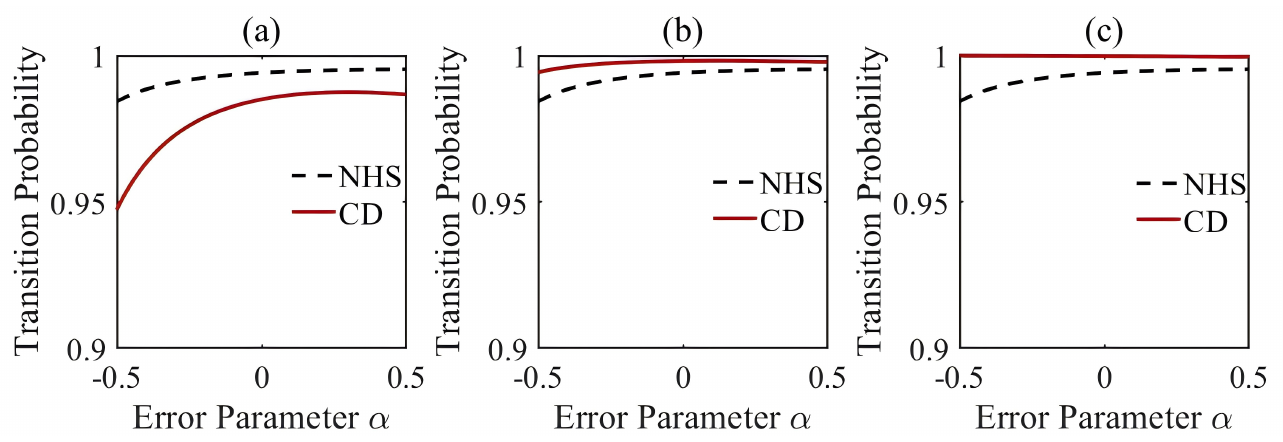}
							\caption{Transition probabilities  versus coupling strength error $\alpha$ (dimensionless
								parameter). In the NHS (black, dashed line), the parameters are $\omega_{c}/\omega_{d}=85, $ $\omega_{m}/\omega_{d}=35$ , $\varepsilon_{m}/\omega_{d}=50$, $g_{m}=1$. In the CD (red, solid line), (a) $g_{m}>\left ( \kappa _{c}+ \kappa _{m} \right )  /  {2}$, $\kappa _{c}=1$, $\kappa _{m} =0.3$, (b) $g_{m}=\left ( \kappa _{c}+ \kappa _{m} \right )  /  {2}$, $\kappa _{c}=1$, $\kappa _{m} =1$, (c) $g_{m}<\left ( \kappa _{c}+ \kappa _{m} \right )  /  {2}$, $\kappa _{c}=2$, $\kappa _{m} =2$. All
								parameters are the same as those in Figs. \ref{fig1} and \ref{fig2}, respectively }\label{fig3}
						\end{center}
					\end{figure*}
					%%%%%%%%%%%%%%%%%%%%%%%%%%%%%%%%%%%%%%%%%%%%%%%%%%%%%%%%%%%%%%
					%%%%%%%%%%%%%%%%%%%%%%%%%%%%%%%%%%%%%%%%%%%%%%%%%%%%%%%%%%%%%%
					%%%%%%%%%%%%%%%%%%%%%%%%%%%%%%%%%%%%%%%%%%%%%%%%%%%%%%%%%%%%%%
					\begin{figure*}[t]%[tpb]  % picture 1
						\begin{center}
							\includegraphics[width=15 cm,angle=0]{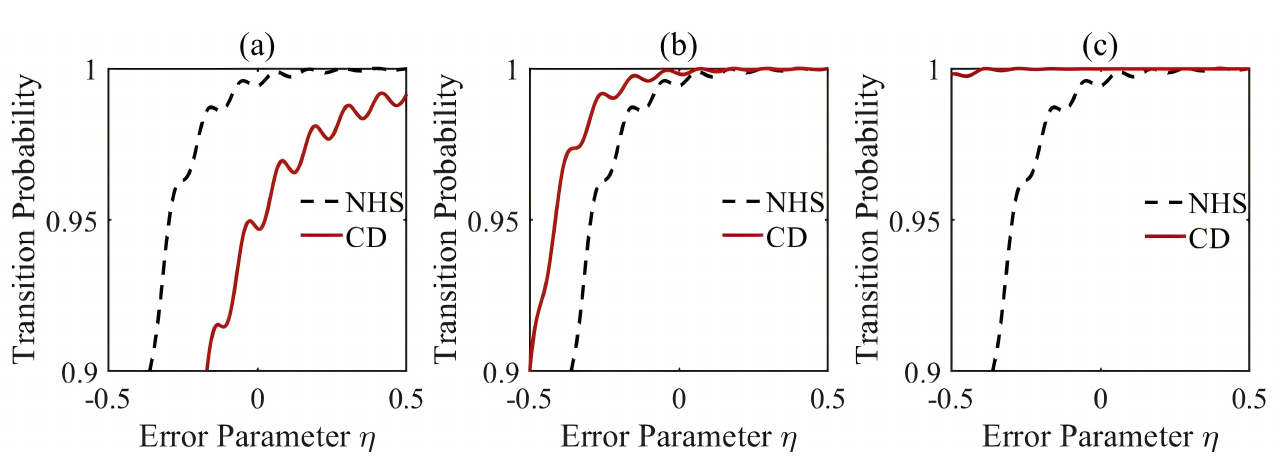}
							\caption{Transition probabilities  versus systematic error $\eta$ (dimensionless
								parameter): In the NHS (black, dashed line), the parameters are $\omega_{c}/\omega_{d}=85, $ $\omega_{m}/\omega_{d}=35$ , $\varepsilon_{m}/\omega_{d}=50$, $g_{m}=1$. In the CD (red, solid line), (a) $g_{m}>\left ( \kappa _{c}+ \kappa _{m} \right )  /  {2}$, $\kappa _{c}=1$, $\kappa _{m} =0.3$, (b) $g_{m}=\left ( \kappa _{c}+ \kappa _{m} \right )  /  {2}$, $\kappa _{c}=1$, $\kappa _{m} =1$, (c) $g_{m}<\left ( \kappa _{c}+ \kappa _{m} \right )  /  {2}$, $\kappa _{c}=2$, $\kappa _{m} =2$. All
								parameters are the same as those in Figs.  \ref{fig1} and \ref{fig2}, respectively}\label{fig4}
						\end{center}
					\end{figure*}
					In CD technique, non-adiabatic transitions can be suppressed by introducing an external driving field $H_{c}(t)$, and the system evolves along the adiabatic path even under non-adiabatic conditions. Assisted by the
					counterdiabatic field, the two-level NH Hamiltonian in the Eq.(\ref{H}) can be rewritten as
						\begin{align}
							H_{tol}=\left(\begin{array}{cc}
								\omega_{c}+i \kappa_{c} & g_{m}+Q(t) \\
								g_{m}-Q(t) & \omega_{m}(t)-i \kappa_{m}
							\end{array}\right) ,
							\end{align}	
				 In Fig. \ref{fig2}, it shows the population evolution of the two-level NH system under different conditions using the CD technique. Overall, initially, the system is mainly in energy level $\left | 0  \right \rangle$  ($P_{0}^{r}=1$  and $P_{1}^{r}=0$). As time goes by, the population $P_{0}^{r}$ of the level $\left | 0  \right \rangle$ gradually decreases, while the population $P_{1}^{r}$ of energy level $\left | 1  \right \rangle$ gradually increases to 1. When the system is in the  broken-$\mathcal{P} \mathcal{T}$-symmetric regime,  it can be observed that the rate of population evolution also increases accordingly as the parameter $\kappa _{c}$ and $\kappa _{m}$ increases, which the population $P_{r}^{1}$ improves at the final time $t=2$, increasing from approximately 98.4$\%$ in Fig. \ref{fig2}(a) to above 99.9$\%$ in Fig. \ref{fig2}(d). Compared with NHS technique, CD technique demonstrates significant advantages in suppressing non-adiabatic coupling and enhancing the evolution speed, thereby achieving more efficient population transfer between microwave photons and magnons.
					
					\section{ Robustness against different experimental errors}
					\label{sec4}
					
					Due to the environmental noise, parameter fluctuations and inaccurate experimental implementation in the actual situation, the parameters of Hamiltonian usually deviate from the ideal scheme, which often leads to the evolution of the system dynamics along an unexpected direction, and ultimately leads to a dramatic decrease in the fidelity of the system target state \cite{1,PhysRevA.98.053413,PhysRevA.103.033110,Xu:22}. In the following, we discuss how the effectiveness of population inversion is affected by several types of experimental errors, including coupling strength error, and systematic error, for the quantum control protocols discussed above.

					\subsection{Coupling strength error}
					We first consider the coupling strength error occurs in the two-level NH system, meaning that the coupling strength $g_{m}$ has an trivial deviation.
					For NHS, to quantify the uncontrolled coupling strength variation, we introduce a dimensionless parameter $\alpha$ to write the coupling strength as
					%%%%%%%%%%%%%%%%%%%%%%%%%%%%%%%%%%%%%%%%%%%%%%%%
					\begin{align}
						g_{m}\rightarrow(1+\alpha)g_{m}
					\end{align}
					%%%%%%%%%%%%%%%%%%%%%%%%%%%%%%%%%%%%%%%%%%%%%%%%%
					For CD technique, the errors of both original Hamiltonian H and additional
					Hamiltonian $H_{c}$ should be considered, i.e., $(g_{m}\pm Q)\rightarrow(1+\alpha)(g_{m}\pm Q)$.
					In  Fig. \ref{fig3}, we compare the accuracies of the two methods by
					plotting the transition probability as a function of error parameter $\alpha$. In the NHS technique, the parameters are $\omega_{c}/\omega_{d}=85, $ $\omega_{m}/\omega_{d}=35$ , $\varepsilon_{m}/\omega_{d}=50$, $g_{m}=1$. By the CD technique, when  the parameters satisfy the relation $g_{m}>\left ( \kappa _{c}+ \kappa _{m} \right )  /  {2}$, we can find that the transition probabilities of CD technique is worse than the  NHS technique. When $g_{m}=\left ( \kappa _{c}+ \kappa _{m} \right )  /  {2} $  and $g_{m}<\left ( \kappa _{c}+ \kappa _{m} \right )  /  {2} $, the CD technique is stable versus variations of $\alpha$ that the fidelity of the transition probability maintains ultrahigh efficiencies above $99.4\%$ in the range
					of $\alpha\in[-0.5,0.5]$. When in the  broken-$\mathcal{P} \mathcal{T}$-symmetric regime, we can conclude that under the condition of fixed parameters of the NHS technique, as the intensity of dissipation $\kappa_m$ increases, the robustness of the CD technique to the coupling strength error is significantly better than that of the NHS technique.
					%%%%%%%%%%%%%%%%%%%%%%%%%%%%%%%%%%%%%%%%%%%%%%%%%%%%%%%%%%%%

					\subsection{Systematic error}
					%%%%%%%%%%%%%%%%%%%%%%%%%%%%%%%%%%%%%%%%%%%%%%%%%%%%%%%%%%%%%%
					\begin{figure*}[t]%[tpb]  % picture 1
						\begin{center}
							\includegraphics[width=18 cm,angle=0]{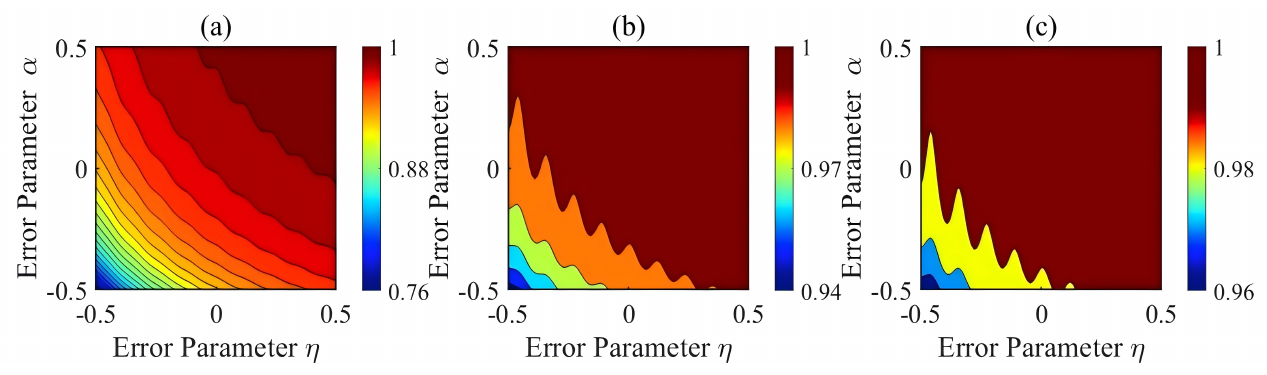}
							\caption{The figures plot the transition probability versus systematic error $\eta$  and  coupling strength error $\alpha$  by NHS.  (a) $g_{m} /\omega_{d}=0.1$, (b) $g_{m} /\omega_{d}=0.6$, (c) $g_{m} /\omega_{d}=1$. All
								parameters are the same as those in Figs.  \ref{fig1} }\label{figTo1}
						\end{center}
					\end{figure*}
					%%%%%%%%%%%%%%%%%%%%%%%%%%%%%%%%%%%%%%%%%%%%%%%%%%%%%%%%%%%%%%
					\begin{figure*}[t]%[tpb]  % picture 1
						\begin{center}
							\includegraphics[width=18 cm,angle=0]{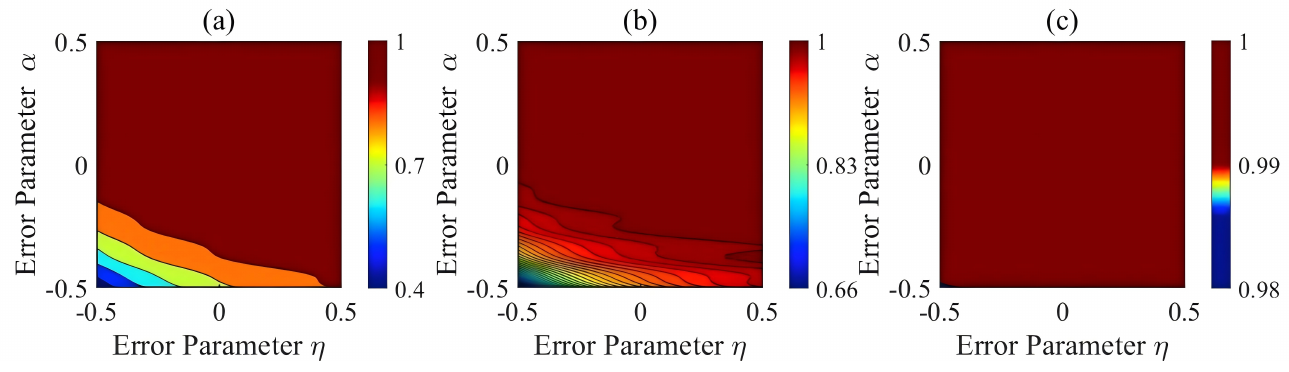}
							\caption{The figures plot the transition probability versus systematic error $\eta$  and  coupling strength error $\alpha$ by CD. (a) $g_{m}>\left ( \kappa _{c}+ \kappa _{m} \right )  /  {2}$, $\kappa _{c}=1$, $\kappa _{m} =0.3$, (b) $g_{m}=\left ( \kappa _{c}+ \kappa _{m} \right )  /  {2}$, $\kappa _{c}=1$, $\kappa _{m} =1$, (c) $g_{m}<\left ( \kappa _{c}+ \kappa _{m} \right )  /  {2}$, $\kappa _{c}=2$, $\kappa _{m} =2$. All
								parameters are the same as those in Fig.  \ref{fig2}}\label{figTo}
						\end{center}
					\end{figure*}
					%%%%%%%%%%%%%%%%%%%%%%%%%%%%%%%%%%%%%%%%%%%%%%%%%%%%%%%%%%%%%%
					%%%%%%%%%%%%%%%%%%%%%%%%%%%%%%%%%%%%%%%%%%%%%%%
					%%%%%%%%%%%%%%%%%%%%%%%%%%%%%%%%%%%%%%%%%%%%%%%%%%%%%%%%%%%%%%
					%%%%%%%%%%%%%%%%%%%%%%%%%%%%%%%%%%%%%%%%%%%%%%%%%%%%%%%%%%%%%%
					%%%%%%%%%%%%%%%%%%%%%%%%%%%%%%%%%%%%%%%%%%%%%%%%%%%%%%%%%%%%%%
					The systematic error originating from imperfection of experimental implementations and conditions will decrease the effectiveness of population
					transfer. The NHS method considers the error of Hamiltonian $H$, and
					the CD method discusses the error of both original Hamiltonian $H$ and the additional Hamiltonian $H_{c}$, i.e., $(H+H_{c})\rightarrow(1+\eta) (H+H_{c})$.
					In Fig. \ref{fig4}, we plot the stability of the transition probability versus the parameter of the systematic error $\eta$. When $g_{m}>\left ( \kappa _{c}+ \kappa _{m} \right )  /  {2}$, the NHS technique is superior to the CD technique. 
					When  $g_{m}<\left ( \kappa _{c}+ \kappa _{m} \right )  /  {2}$ , the CD technique is stable versus variations of $\eta$  that  the fidelity of the transition probability maintains ultrahigh efficiencies
					above $99.73\% $ in the range of the systematic error  $\eta \in[-0.5,0.5]$. For $g_{m}=\left ( \kappa _{c}+ \kappa _{m} \right )  /  {2}$, the efficiencies of the CD and NHS technologies keep the probability
					at ultrahigh efficiencies above $99.9\%$ when $\eta\in[0.2,0.5]$, while CD technique is superior to the NHS when $\eta\in[-0.5,0.2]$. Overall, the CD technique is superior to another coherent control method with respect to the systematic error.
					
					Furthermore, we examine how the accuracy of transition probabilities is simultaneously affected by two experimental errors for the two coherent
					control techniques. First, we plot
					the transition probability as a function of  the systematic error $\eta$  and the rabi frequency error $\alpha$ by NHS in Fig. \ref{figTo1}. The transition probabilities  steadily
					improve as $\alpha$ increase. Then, we plot how the fidelity
					is affected by the systematic error $\eta$ and the coupling strength error $\alpha$ by CD in Fig. \ref{figTo}.
					The CD method
					is the clear winner, followed by the NHS technique.
					Compare to the NHS method, the CD possess a much higher magnitude of the transition probability and feature a broad
					range of high efficiencies with the fidelity being $99.9\%$, showing its ultrahigh robustness
					against the two experimental errors.
					
					%%%%%%%%%%%%%%%%%%%%%%%%%%%%%%%%%%%%%%%%%%%%%%%%%%%%%%%%%%%%%%%
					\section{Summary}
					\label{sec5}
					In summary, we use CD and the Floquet engineering to realize the robust and fast state transfer in the dissipation cavity magnon-polaritons NH system.
					The dissipative cavity-magnon system consists of a magnon in a small
					yttrium iron garnet sphere and a three-dimensional rectangular microwave cavity, which can be regarded as a two-level NH system.
					Based on the coherent control techniques, including NHS and CD, a nearly perfect relative population transfer is achieved from the microwave photons to the magnons.
					By comparing the sensitivities and robustness of the NHS and CD techniques to coupling strength and systematic errors, we find that the CD method  performs better in the presence of these experimental errors. In the  broken-$\mathcal{P} \mathcal{T}$-symmetric regime, the CD method maintains the transition probability at
					ultrahigh the fidelity above 99.9\% over broad ranges of the above two errors, and this robustness continues to enhance as the gain rate increases. When the two-level NH system is in the broken-$\mathcal{P} \mathcal{T}$-symmetric regime, the advantages of CD technique are more significant than the NHS technique. It can better maintain high-fidelity state transitions and demonstrate stronger robustness.
					This makes CD technique is a promising technique for coherent control of
					two-level NH quantum systems in cavity electromagnonics.

					\begin{acknowledgements}
						This work was supported by the National Natural Science Foundation of China (Grant No. 62471001, No. 12475009, No. 12075001, and No. 12175001), Natural Science Research Project in Universities of Anhui Province (Grant No. 2024AH050068),
						Anhui Provincial Key Research and Development Plan (Grant No. 2022b13020004), and Anhui Provincial Natural Science Foundation (Grant No. 1508085QF139).
					\end{acknowledgements}

\end{document}